\documentclass[twocolumn,english,prx,superscriptaddress]{revtex4-2}
\usepackage{color}
\usepackage[utf8]{inputenc}
\usepackage{graphicx}
\usepackage[T1]{fontenc}
\usepackage{float}
\usepackage{natbib}
\usepackage{textcomp}
\usepackage{amsmath}
\usepackage{amssymb}
\usepackage{graphicx}
\usepackage{esint}
\usepackage{natbib}
\usepackage{xcolor}
\usepackage{multirow}
\usepackage{ulem}
\usepackage{comment}
\graphicspath{{Figures/}}
\usepackage{dsfont}

\begin{document}
\title{Scaled tight binding model for a two dimensional electron gas at the (001) LaAlO$_3$/SrTiO$_3$ interface. }
 
\author{Paweł Wójcik}
\email{pawelwojcik@agh.edu.pl}
\affiliation{$^1$AGH University of Krakow, Faculty of Physics and
Applied Computer Science, al. Mickiewicza 30, 30-059 Kraków, Poland}

\author{Roberta Citro}
\email{rocitro@unisa.it}
\affiliation{Physics Department ‘E. R. Caianiello’ and CNR-SPIN, Università degli Studi di Salerno, Fisciano, (SA), Italy}

\author{Bartłomiej Szafran}
\email{bszafran@agh.edu.pl}
\affiliation{$^1$AGH University of Krakow, Faculty of Physics and
Applied Computer Science, al. Mickiewicza 30, 30-059 Kraków, Poland}


\begin{abstract}
The progress in the fabrication of nanoscale systems based on the two-dimensional electron gas at the interface between LaAlO$_3$ and SrTiO$_3$ (LAO/STO) has created an increased demand for simulations of these nanostructures, which typically range in size from tens to hundreds of nanometers. Due to the low lattice constant of LAO/STO, approximately 0.394 nm, these calculations become extremely time-consuming. Here, we present a scaled tight-binding approximation defined on a mesh with size that can be several times larger than in the ordinary approach. The scaled model is analyzed within the context of quantum transport simulations and electronic structure calculations. Our findings demonstrate that the scaled model closely aligns with the ordinary one up to a scaling factor of 8. These results pave the way for more efficient simulations of LAO/STO nanostructures with realistic sizes relevant to experimental applications.
\end{abstract}

\maketitle

\section{Introduction}
\label{sec:intro}
Transition metal oxide interfaces host a rich spectrum of functional properties~\cite{Chen2024}, making them a promising material platform in many fields, including electronics~\cite{Ohtomo2004}, spintronics~\cite{Lesne2016,Vaz2019,Trama2022,Trama2022_2}, and quantum computing~\cite{barthelemy2021}. Among the perovskite oxide family, a two-dimensional electron gas (2DEG) at the interface between LaAlO$_3$ and SrTiO$_3$ (LAO/STO) is one of the best-known and most extensively investigated. It is characterized by a unique combination of high mobility~\cite{Ohtomo2004}, large spin-orbit coupling (SOC)~\cite{Diez2015,Rout2017,Caviglia2010,Shalom2010,Yin2020,Singh2017,Hurand2015}, gate-tunable superconductivity~\cite{Reyren2007, joshua2012universal,maniv2015strong,
Biscaras, Monteiro2019, Monteiro2017, Zegrodnik2020, Wojcik_2021} or magnetic ordering~\cite{Karen2012,Li2011,Brinkman2007,Dikin2011,Bert2011}. The electronic structure of the LAO/STO interface is governed by the $d$-orbitals of Ti~\cite{Popovic2008, Pavlenko2012, Pentcheva2006}, whose properties exhibit exceptional sensitivity to reconstruction by external disturbances. Both static and dynamic manipulation of the $d$-orbital degrees of freedom offer precise control over the functionalities of the 2DEG through external electric fields, which determine its potential applications.

The significant progress in epitaxial growth methods of transition metal oxides currently allows for the fabrication of 2DEGs with specific functional properties, such as interfacial 2D-ferromagnetism (FM)~\cite{Stornaiuolo2018} and ferroelectricity (FE)~\cite{Gastiasoro2020,Kanasugi2018,Kanasugi2019,Honig2013}. One of their advantage is the ability to use local top, split, or side gates to define device of desired geometry~\cite{joshua2012universal,Biscaras, Monteiro2017,Caviglia2010,Hosoda2013,Goswami2015, Hurand2015,Thierschmann2018, Jouan2020, Settino2021}. The susceptibility of LAO/STO 2DEG to electrostatic gating, comparable to semiconductor materials, has led to significant advances in oxide-2DEG nanotechnology, which now allows for the realization of complex structures down to a scale of tens of nanometers. In particular, the quantization of conductance in ballistic quantum point contact (QPCs) formed by the electrostatic gating has been recently demonstrated experimentally in Ref.~\cite{Jouan2020}. 

Another key property of the LAO/STO interface is related with the $d$-orbital character of the electrons confined at the interface, which implies that direct and indirect decoherence stemming from interaction with the nuclear bath can be significantly mitigated. Note that the decoherence is proportional to the square of the wave function at the nuclei's position, which tends to zero for $3d$ electrons~\cite{Loss2002}. This property, combined with the strong spin-orbit interaction, has initiated ongoing research on LAO/STO-based quantum dots as potential spin qubits with the inherent scalability of 2D systems~\cite{Guenevere2017,Jespersen2020}. The first experimental realization of electrostatically defined LAO/STO quantum dots has already been reported in Ref.~\cite{Jespersen2020}, yielding Coulomb diamonds~\cite{Jespersen2020}. Remarkably, some of these experiments provide experimental evidence of the attractive interaction between electrons in QDs, which could shed light on the pairing mechanism in the LAO/STO interface~\cite{Guenevere2017}.

Along the experimental breakthroughs in the field of the LAO/STO interfaces, theoretical analyses of the emerging experiments have been reported, both in the case of transport through the QPC~\cite{Settino2021} and in the electrical spin manipulation in single and double quantum dots~\cite{Szafran2024,Szafran2024_2}. The results of the latter, still awaiting experimental verification, are particularly promising, predicting spin flip time much shorter than those extracted for QDs based on semiconductor materials. Note, however, that the simulations of realistic quantum devices based on LAO/STO remains a challenging problem, as the tight-binding Hamiltonian for the interface is defined on a square lattice with a size of grid spacing $a = 0.395$ nm. This makes simulations of devices with realistic sizes of a few hundred of nanometers extremely time-consuming or even impossible.

Here, we present a scaled tight-binding approximation for describing the (001) LAO/STO 2DEG, defined on a square lattice of arbitrary size. The proposed model is analyzed in the context of quantum transport simulations and calculations of the electronic structure of single and two-electron quantum dots.

The manuscript is organized as follows: In Sec. II, we present the tight-binding approximation in the wave vector space for the 2DEG at the (001) LAO/STO interface along with its scaled version, defined in the real space. Sec. III contains the analysis of the scaled model in quantum transport simulations and electronic spectrum calculations. Finally, the summary is included in Sec. IV.

\section{Theoretical model}
\label{sec:theory}
\subsection{TBA in wave vector space}
The electronic structure of the 2DEG at the (001)-oriented LAO/STO interface is described by the Ti $t_{2g}$ orbitals  ($d_{xy}$, $d_{yz}$, $d_{xz}$) coupled through the $2p$ states of oxygen, localized on the square 
lattice~\cite{Popovic2008,Pavlenko2012, Pentcheva2006, Diez2015}. Although in the bulk STO, all the $d$-electronic states are degenerate at the $\Gamma$ point of the Brillouin zone, the angle-resolved photoemission spectroscopy revealed that the energy of the $d_{xy}$ state at the LAO/STO interface is shifted by $47$~meV with respect to the energy of the $d_{yz}$, $d_{xz}$ bands as a results of the confinement at the interface~\cite{maniv2015strong}. 
The Hamiltonian of (001) LAO/STO 2DEG is given by
\begin{equation}
    \hat{H}_{\mathbf{k}}=\sum _{\mathbf{k}} \hat{C}^\dagger_{\mathbf{k}} ( \hat{H}_{0}+\hat{H}_{RSO}+\hat{H}_{SO}+\hat{H}_B ) \hat{C}_{\mathbf{k}},
    \label{eq:Hamiltonian_k_space}
\end{equation}
where $\hat{C}_{\mathbf{k}}=(\hat{c}_{\mathbf{k},xy}^{\uparrow}, \hat{c}_{\mathbf{k},xy}^{\downarrow}, \hat{c}_{\mathbf{k},xz}^{\uparrow}, \hat{c}_{\mathbf{k},xz}^{\downarrow}, \hat{c}_{\mathbf{k},yz}^{\uparrow}, \hat{c}_{\mathbf{k},yz}^{\downarrow})^{T}$ corresponds to the vector of anihilation operators for electrons with spin $\sigma=\uparrow,\downarrow$ on the orbital $d_{xy},d_{xz},d_{yz}$, in the state $\mathbf{k}$.

In the tight binding approximation, the kinetic energy $\hat{H}_0$ in Eq.~(\ref{eq:Hamiltonian_k_space}) can be expressed as
\begin{equation}
\hat{H}_{0}=
\left(
\begin{array}{ccc}
 \epsilon^{xy}_{\mathbf{k}} & 0 & 0\\
 0 & \epsilon^{xz}_{\mathbf{k}} &  \epsilon^h_{\mathbf{k}} \\
 0 & \epsilon^h_{\mathbf{k}}  &  \epsilon^{yz}_{\mathbf{k}}
\end{array} \right) \otimes \hat {\sigma} _0\;,
\end{equation}
where
\begin{equation}
\begin{split}
    \epsilon^{xy}_{\mathbf{k}}&=4t_l-2t_l\cos{k_xa}-2t_l\cos{k_ya}-\Delta_E,\\
    \epsilon^{xz}_{\mathbf{k}}&=2t_l+2t_h-2t_l\cos{k_xa}-2t_h\cos{k_ya},\\
    \epsilon^{yz}_{\mathbf{k}}&=2t_l+2t_h-2t_h\cos{k_xa}-2t_l\cos{k_ya},\\
    \epsilon^h_{\mathbf{k}}&=2t_d\sin{k_xa}\sin{k_ya}.
\end{split}
\label{eq:H0}
\end{equation}
The latter term is responsible for the hybridization between the $d_{xz}$ and $d_{yz}$ bands, with the strength determined by the parameter $t_d$. The remaining parameters, $t_l$ and $t_h$, are the hopping energies for the light and heavy bands, respectively, and $a$ is a lattice constant $a=0.394$~nm. The tight-binding parameters are reported in Ref. \onlinecite{maniv2015strong} and are taken as: $t_l=875\;$meV, $t_h=40\;$meV, $t_d=40\;$meV, $\Delta_E=47\;$meV.

The SO coupling in the (001) LAO/STO-based 2DEG can be described by  two components: the atomic and the Rashba part. The atomic part is an effect of the atomic $\mathbf{L} \cdot \mathbf{S}$ interaction and is given by~\cite{Khalsa2013}
\begin{equation}
\hat{H}_{SO}= \frac{\Delta_{SO}}{3}
\left(
\begin{array}{ccc}
0 & i \sigma _x & -i \sigma _y\\
-i \sigma _x & 0 & i \sigma _z \\
i \sigma _y & -i \sigma _z & 0
\end{array} \right) \;,
\label{eq:hso}
\end{equation}
where $\Delta _{SO}$ determines the strength of the atomic spin-orbit energy and $\sigma _{x},\sigma_{y},\sigma_{z}$ are the Pauli matrices. \\
The Rashba-like spin-orbit term $\hat{H}_{RSO}$, in turn, results from the the mirror symmetry breaking at the interface and takes the form
\begin{equation}
\hat{H}_{RSO}= \Delta_{RSO}
\left(
\begin{array}{ccc}
0 & i \sin{k_ya} & i \sin{k_xa}\\
-i \sin{k_ya} & 0 & 0 \\
-i \sin{k_xa} & 0 & 0
\end{array} \right) \otimes \hat {\sigma} _0\;,
\label{eq:rso}
\end{equation}
where $\Delta _{RSO}$ determines the energy of the Rashba SO coupling. \\

Finally, the coupling of the external magnetic field to the spin and orbital momentum of electrons is taken into account by the Hamiltonian
\begin{equation}
\hat{H}_B=\mu_B(\mathbf{L}\otimes \sigma_0+g\mathds{1}_{3\times 3} \otimes \mathbf{S})\cdot \mathbf{B}/\hbar,
\label{eq:Hb}
\end{equation}  
where $\mu_B$ is the Bohr magneton, $g$ is the Land\'e factor, $\mathbf{S}=\hbar \pmb{\sigma}/2$ with $\pmb{\sigma}=(\sigma_x,\sigma_y,\sigma_z)$ and $\mathbf{L}=(L_x,L_y,L_z)$ with
\begin{equation}
\begin{split}
 L_x&= \left ( 
 \begin{array}{ccc}
  0 & i & 0 \\
  -i & 0 & 0 \\
  0 & 0 & 0 
 \end{array}
 \right ), 
 L_y= \left ( 
 \begin{array}{ccc}
  0 & 0 & -i \\
  0 & 0 & 0 \\
  i & 0 & 0 
 \end{array}
 \right ), 
 L_z= \left ( 
 \begin{array}{ccc}
  0 & 0 & 0 \\
  0 & 0 & i \\
  0 & -i & 0 
 \end{array}
 \right ).
 \end{split}
\end{equation}

\begin{figure*}[!t]
\begin{tabular}{cccc}
\includegraphics[width=.55\columnwidth]{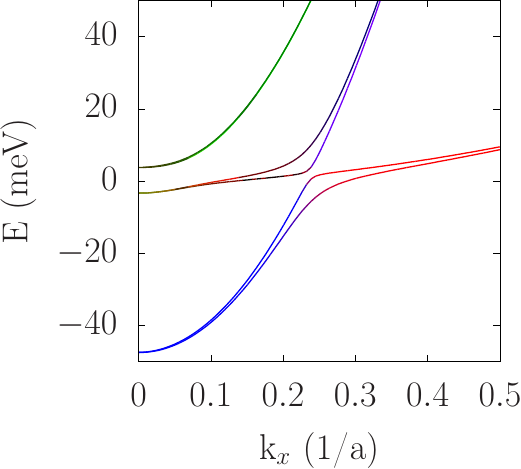} \put(-95,110){(a)}&\includegraphics[width=0.55\columnwidth]{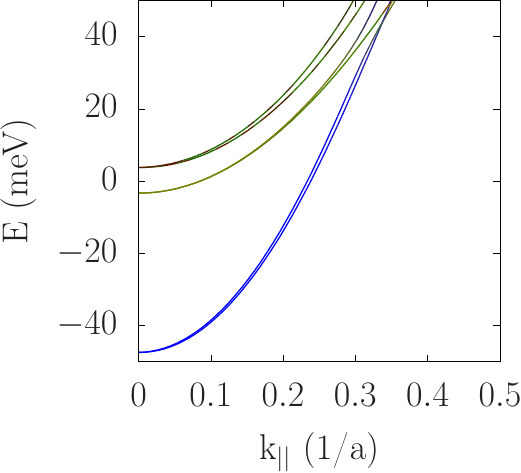}\put(-95,110){(b)}&\includegraphics[width=0.56\columnwidth]{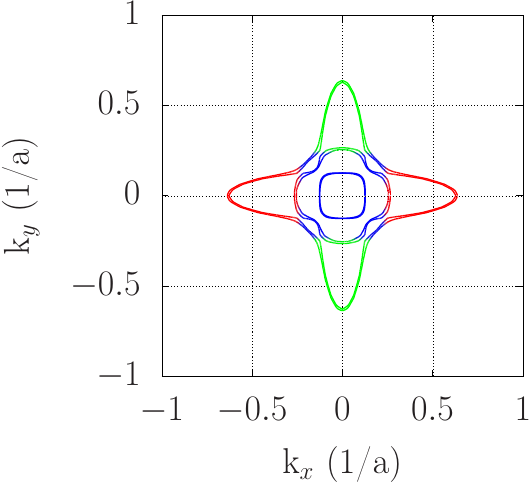}\put(-93,110){(c)}
\end{tabular}
\caption{Dispersion relations $E(k)$ determined from the TBA  defined in the wave  vector space along (a) $k_x$ and (b) $k_{||}=(k,k)$. (c) Fermi surface for $E_F=15$~meV. The colors indicate the type of orbital: $d_{xy}$ is marked by blue, $d_{xz}$ by green and $d_{yz}$ by red in the rgb system.}
\label{fig1}
\end{figure*}

The dispersion relation $E(k)$ for the SO coupling parameters $\Delta_{SO}=10$~meV, $\Delta_{RSO}=20$~meV corresponding to that measured experimentally~\cite{Caviglia2010,Yin2020} are presented in 
Fig.~\ref{fig1}.\\

\subsection{Scaled TBA in real space}
Simulations of nanostructures based on the LAO/STO 2DEG require the formulation of the Hamiltonian (\ref{eq:Hamiltonian_k_space}) in real space, as was done in one of our previous studies~\cite{Szafran2024}. Note, however, that the real-space TBA of (001) LAO/STO is defined on a square lattice with the grid constant being a fraction of a nanometer ($a=0.395$~nm). This significantly restricts the size of systems that can be considered within this model, making simulations of realistically sized devices, comparable to those experimentally fabricated, extremely time-consuming.

Here, we present a low-energy scaled Hamiltonian for the (001) LAO/STO 2DEG, defined on a square lattice with a size $dx$ that can be several times larger than the lattice constant $a$. Within this model, the Hamiltonian of the LAO/STO interface is given by
\begin{eqnarray}
    \hat{H}&=&\sum _{\mu,\nu} \hat{C}^\dagger_{\mu, \nu} ( \hat{H}^{0}+\hat{H}_{SO}+\hat{H}_B ) \hat{C}_{\mu, \nu}+ 
    \nonumber \\
    &&\sum _{\mu,\nu} \hat{C}^\dagger_{\mu+1, \nu} \hat{H}^{x} \hat{C}_{\mu, \nu}  + \sum _{\mu,\nu} \hat{C}^\dagger_{\mu, \nu+1} \hat{H}^{y} \hat{C}_{\mu, \nu} + \label{eq:Hamiltonian_real} \\ 
    &+& \sum _{\mu,\nu} \hat{C}^\dagger_{\mu+1, \nu-1} \hat{H}_{mix} \hat{C}_{\mu, \nu}-
    \sum _{\mu,\nu} \hat{C}^\dagger_{\mu+1, \nu+1} \hat{H}_{mix} \hat{C}_{\mu, \nu} +h.c., \nonumber
\end{eqnarray}
where $\hat{C}_{\mu, \nu}=(\hat{c}_{\mu, \nu,xy}^{\uparrow}, \hat{c}_{\mu, \nu, xy}^{\downarrow}, \hat{c}_{\mu, \nu, xz}^{\uparrow}, \hat{c}_{\mu, \nu, xz}^{\downarrow}, \hat{c}_{\mu, \nu, yz}^{\uparrow}, \hat{c}_{\mu, \nu, yz}^{\downarrow})^{T}$ corresponds to the vector of anihilation operators of electron with spin $\sigma=\uparrow,\downarrow$ on the orbital $d_{xy}, d_{xz}, d_{yz}$ and the position on the real-space lattice determined by the indexes $( \mu,\nu )$.

The on-site energies and the hopping amplitudes are defined by the operators 
\begin{eqnarray}
\hat{H}^{0}&=&
\left(
\begin{array}{ccc}
 4t^\prime_l-\Delta _E & 0 & 0\\
 0 & 2t^\prime_l+2t^\prime_h & 0 \\
 0 & 0  &  2t^\prime_l+2t^\prime_h
\end{array} \right) \otimes \hat {\sigma} _0 \nonumber \\
&+&
\left(
\begin{array}{ccc}
 V_{\mu,\nu} & 0 & 0\\
 0 & V_{\mu,\nu} & 0 \\
 0 & 0  &  V_{\mu,\nu}
\end{array} \right) \otimes \hat {\sigma} _0 ,
\label{eq:Hamiltonian_real_H0}
\end{eqnarray}
\vspace{0.5cm}
\begin{equation}
\hat{H}^{x}=
\left(
\begin{array}{ccc}
 -t^\prime_l & 0 & 0\\
 0 & -t^\prime_l & 0 \\
 0 & 0  &  -t^\prime_h
\end{array} \right) \otimes \hat {\sigma} _0
+
\frac{\Delta^\prime_{RSO}}{2}
\left(
\begin{array}{ccc}
 0 & 0 & -1 \\
 0 & 0 & 0 \\
 1 & 0  &  0
\end{array} \right) \otimes \hat {\sigma} _0\;,
\label{eq:real_space_Hx}
\end{equation}
\begin{equation}
\hat{H}^{y}=
\left(
\begin{array}{ccc}
 -t^\prime_l & 0 & 0\\
 0 & -t^\prime_h & 0 \\
 0 & 0  &  -t^\prime_l
\end{array} \right) \otimes \hat {\sigma} _0
+
\frac{\Delta^\prime_{RSO}}{2}
\left(
\begin{array}{ccc}
 0 & -1 & 0 \\
 1 & 0 & 0 \\
 0 & 0  &  0
\end{array} \right) \otimes \hat {\sigma} _0,
\label{eq:real_space_Hy}
\end{equation}
and 
\begin{equation}
\hat{H}_{mix}= \frac{t^\prime_d}{2}
\left(
\begin{array}{ccc}
 0 & 0 & 0 \\
 0 & 0 & 1 \\
 0 & 1  &  0
\end{array} \right) \otimes \hat {\sigma} _0 ,
\label{eq:real_space_hybrid}
\end{equation}
where the prime index denotes parameters scaled by the lattice size: $t^\prime_l=t_l(a/dx)^2$, $t^\prime_h=t_h(a/dx)^2$, $t^\prime_d=t_d(a/dx)^2$ and $\Delta^\prime_{RSO}=\Delta_{RSO}(a/dx)$ with the scaling factor defined by the ratio $s=dx/a$. In Eq.~(\ref{eq:Hamiltonian_real}), $\hat{H}_{SO}$ and $\hat{H}_B$ remains unscaled and have the same form as in the wave vector space formulation, given by Eqs.~(\ref{eq:hso}) and (\ref{eq:Hb}).

\begin{figure}[!h]
\begin{tabular}{cc}
\includegraphics[width=.6\columnwidth]{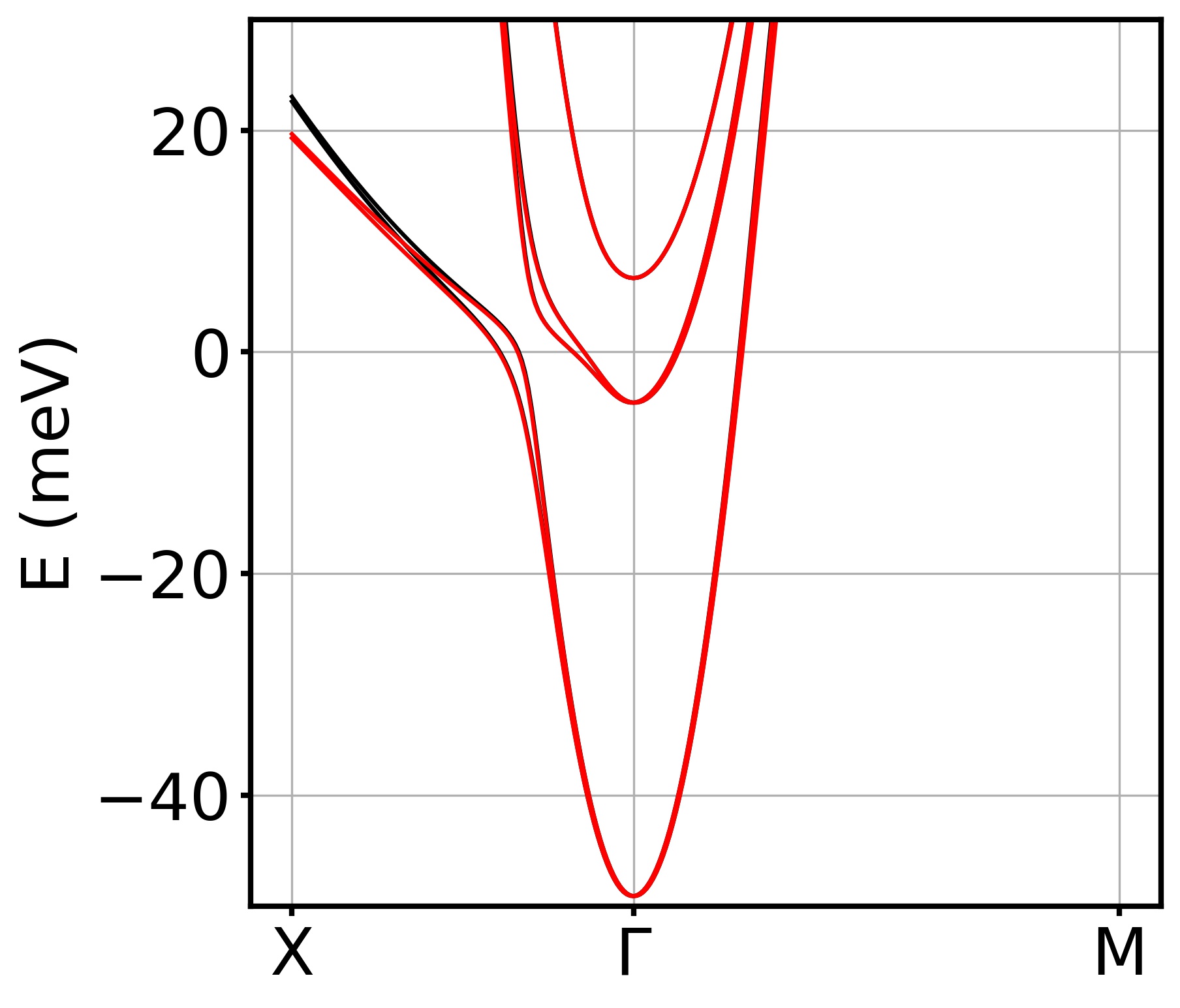} \put(-20,110){(a)} \vspace{0.3cm} \\
\includegraphics[width=0.6\columnwidth]{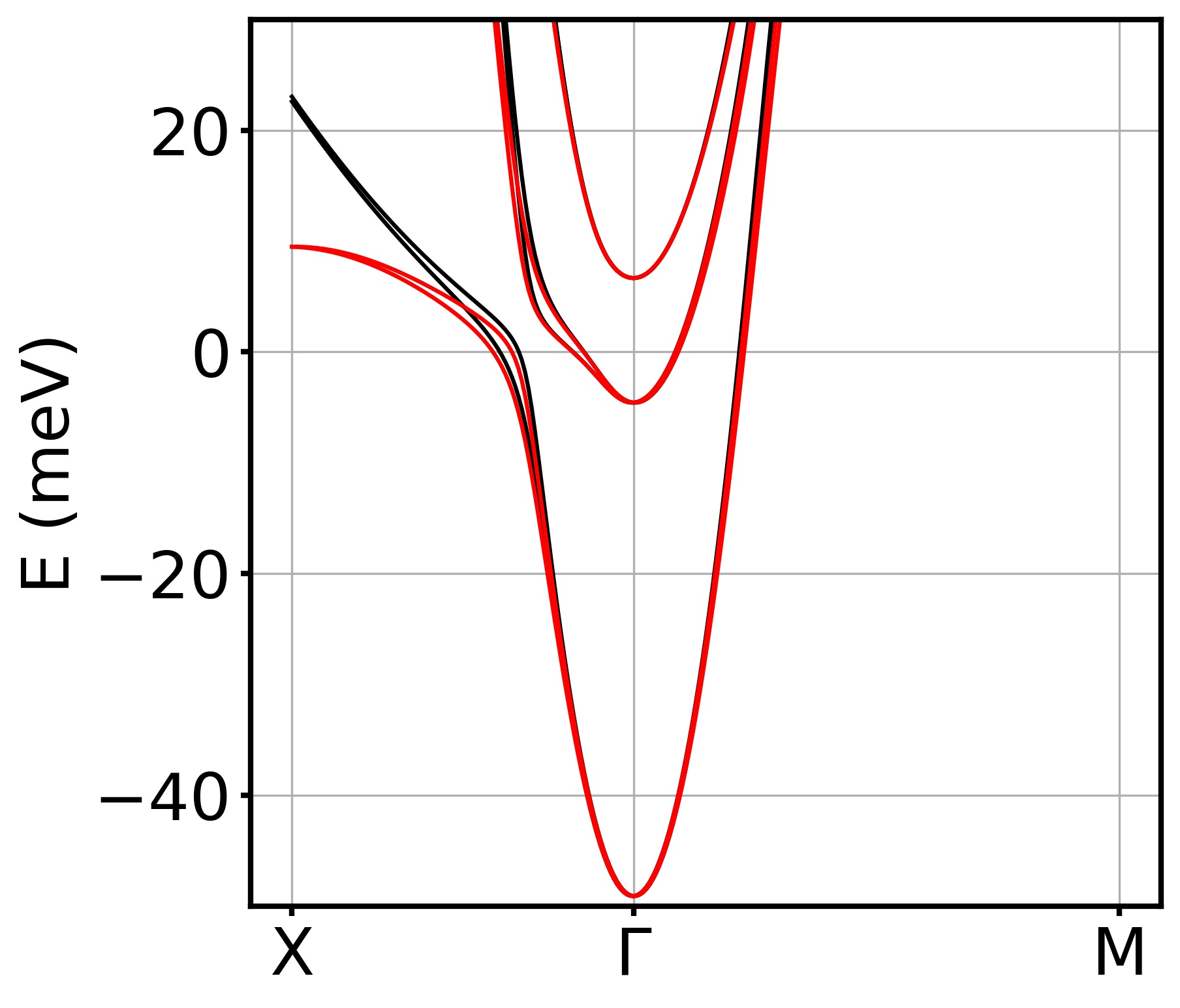}\put(-20,110){(b)} 
\end{tabular}
\caption{Electronic structure $E(k)$ for the (001) LAO/STO interface determined from the TBA (black) and based on the scaled Hamiltonian (red) with the scaling (a) $s=2$ and (b) $s=4$.}
\label{fig2}
\end{figure}
In Fig.~\ref{fig2}, we present a comparison of the electronic structure determined from the TBA in wave vector space (black) and based on the scaled Hamiltonian defined by Eq.~(\ref{eq:Hamiltonian_real}) (red), with the scaling factor (a) $s=2$ and  (b) $s=4$. Our model with the scaling reproduces the electronic structure to a great extent, especially at low energy limit, near the edges of the bands.

\section{Results}
\label{sec:results}
Below, we will explore the capabilities of our model in two key areas: (A) quantum transport simulations, and (B) the determination of single- and two-electron energy spectra of a single and double quantum dots. 

\subsection{Quantum transport simulations}
\begin{figure}[!h]
\includegraphics[width=.8\columnwidth]{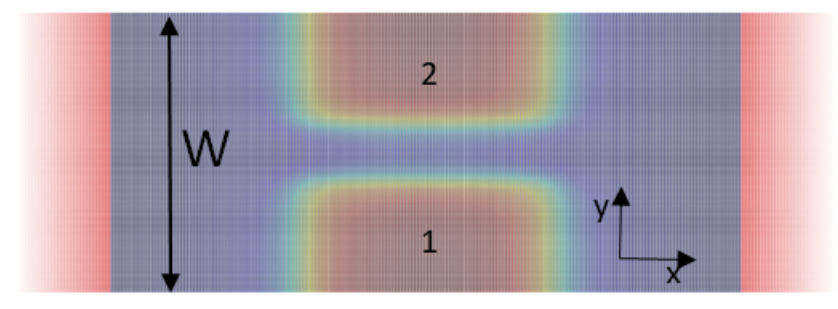} 
\caption{Schematic illustration of the QPC based on the (001) LAO/STO 2DEG.}
\label{fig:QPC}
\end{figure}

Let us first consider the LAO/STO nanowire of width $W$ with a QPC, modeled by the potential~\cite{Ferry}
\begin{eqnarray}
    V_{QPC}(x,y) = &V_g& \big [  f(x-l_1,y-b_1)+f(x-l_1,t_1-y) \nonumber \\
    &+& f(r_1-x,y-b_1)+f(r_1-x,t_1-y) \nonumber \\
    &-& f(x-l_2,y-b_2)+f(x-l_2,t_2-y)  \nonumber \\
    &+& f(r_2-x,y-b_2)+f(r_2-x,t_2-y) \big ] \nonumber \\
\end{eqnarray}
where 
\begin{equation}
f(x,y)=\frac{1}{2\pi}\arctan \left ( \frac{xy}{d\sqrt{x^2+y^2+d^2}}\right )
\end{equation}
and $r,l,t,b$ represent the positions of the right, left, top, and bottom edges of the side electrodes, indexed by $1$ and $2$, respectively. The schematic illustration of the structure is presented in Fig.~\ref{fig:QPC}. Note that a similar device was recently investigated experimentally ~\cite{Jouan2020}, with a minimal constriction size of approximately $25$~nm. 

The calculations were performed using the KWANT package~\cite{Groth2014} within the scaled model (see Sec. II.B), defined on a lattice with a spacing of $dx$.
\begin{figure}[!h]
\begin{tabular}{cc}
\includegraphics[width=.47\columnwidth]{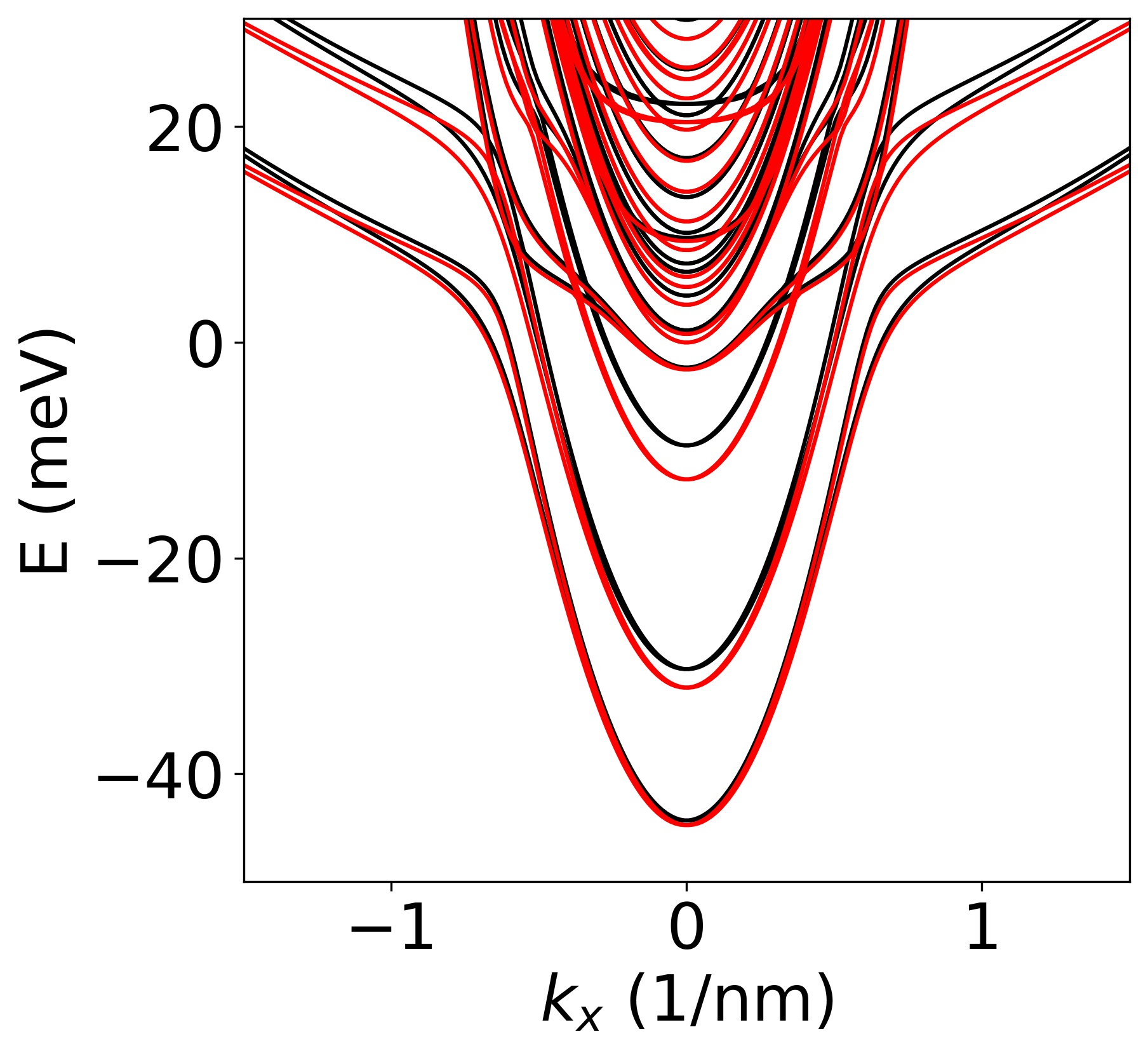} \put(-87,25){(a)}& \includegraphics[width=0.47\columnwidth]{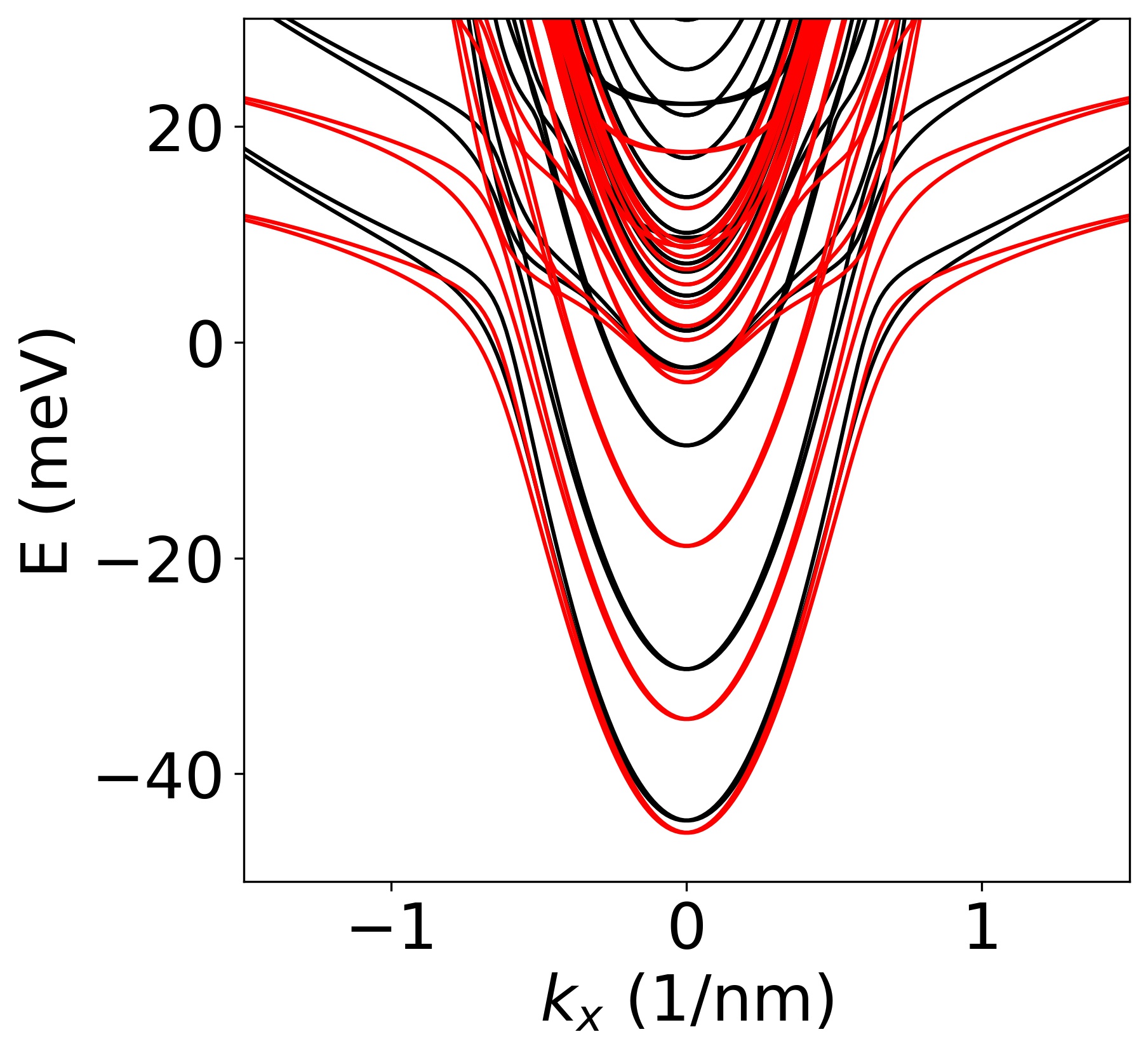}\put(-87,25){(b)} \\
\includegraphics[width=.47\columnwidth]{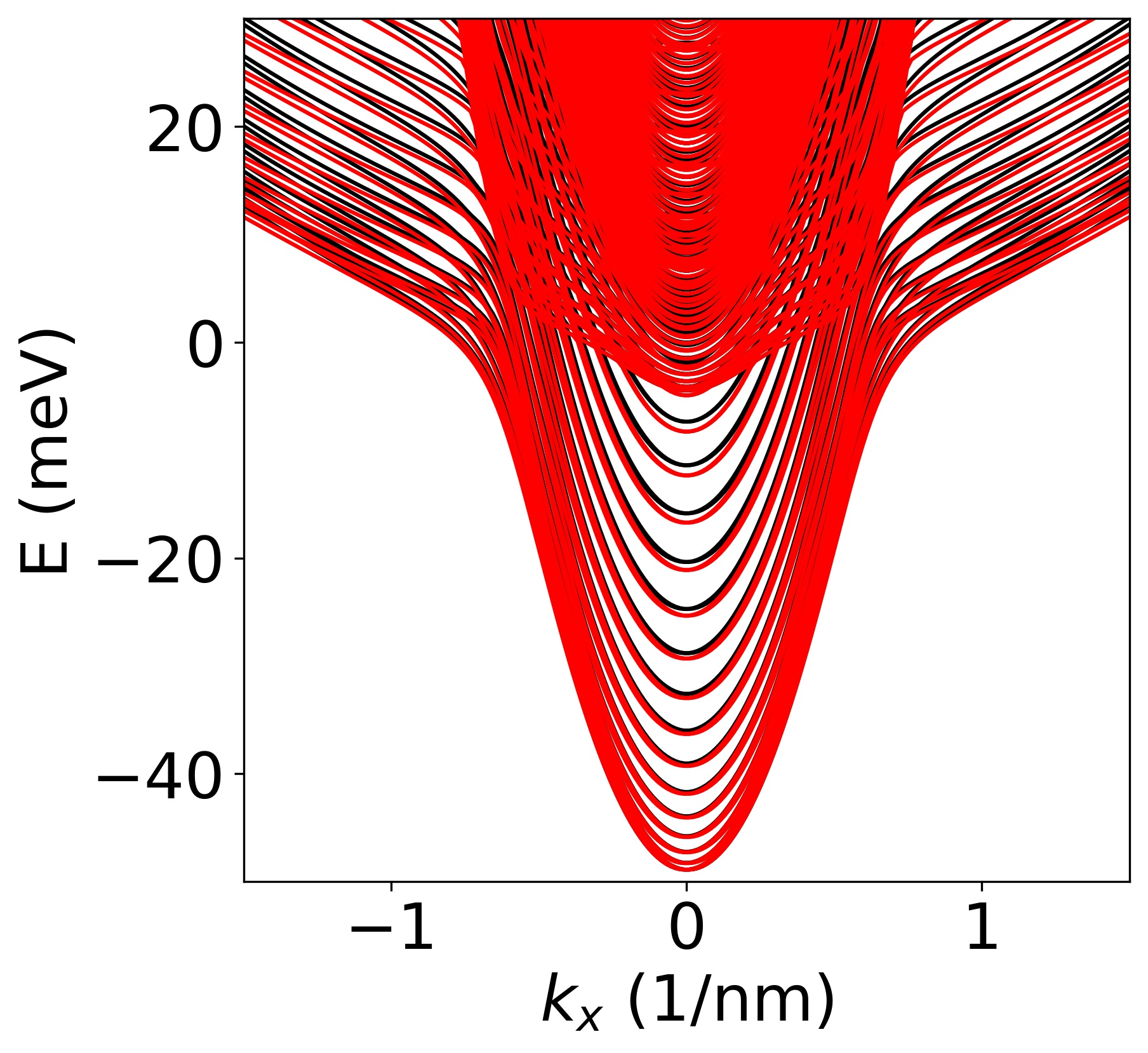} \put(-87,25){(c)}& \includegraphics[width=0.47\columnwidth]{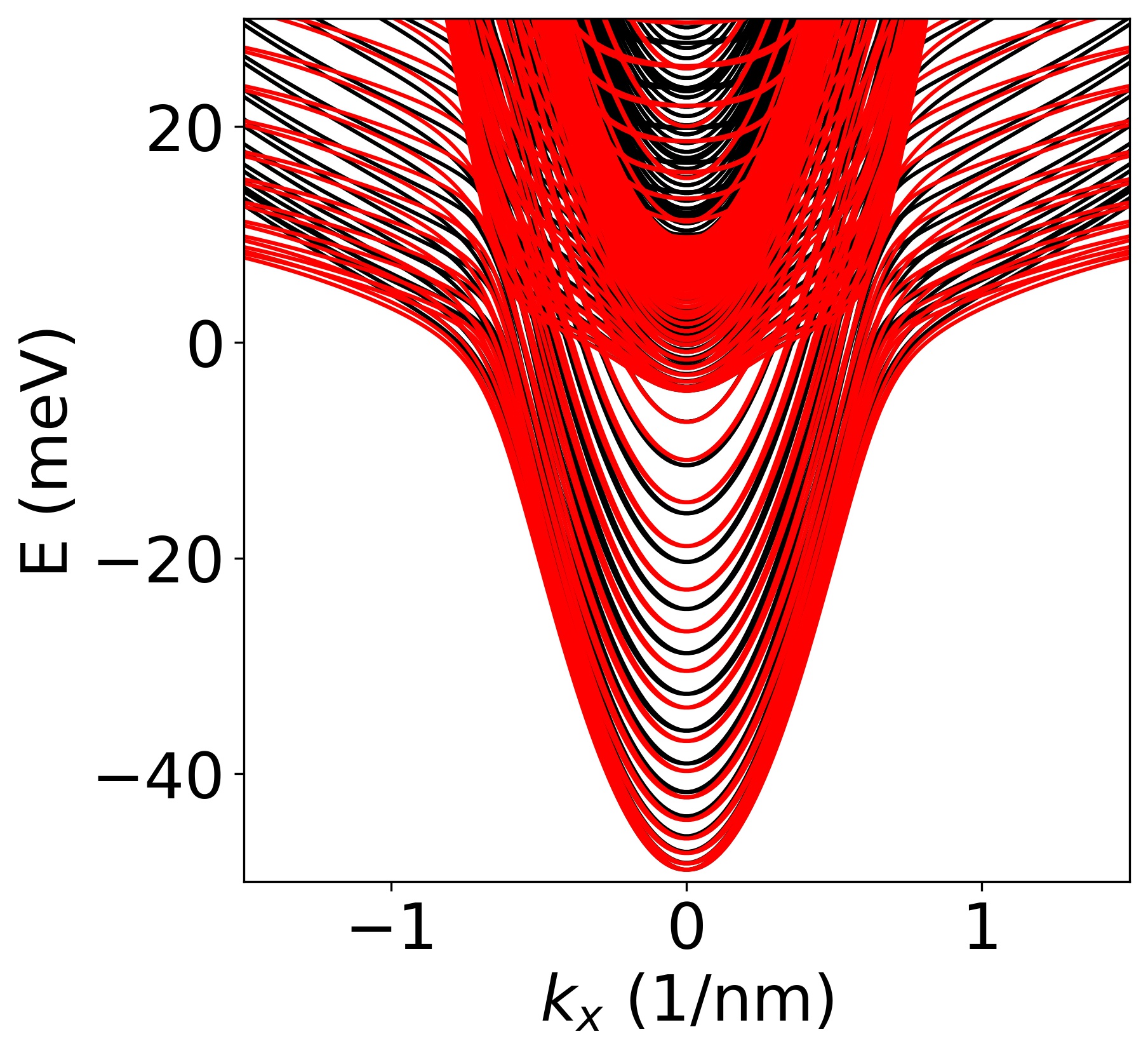}\put(-87,25){(d)}
\end{tabular}
\caption{Electronic spectrum of the LAO/STO nanowire with the width $W=15.6$~nm (a,b) and $W=78$~nm (c,d), determined within the scaled model (red) with the scaling (a,c) $s=2$ and (c,d) $s=4$. Results from the ordinary TBA, defined on a lattice with size $a$, are plotted in black.}
\label{fig:Ek}
\end{figure} 
\begin{figure}[!h]
\includegraphics[width=.6\columnwidth]{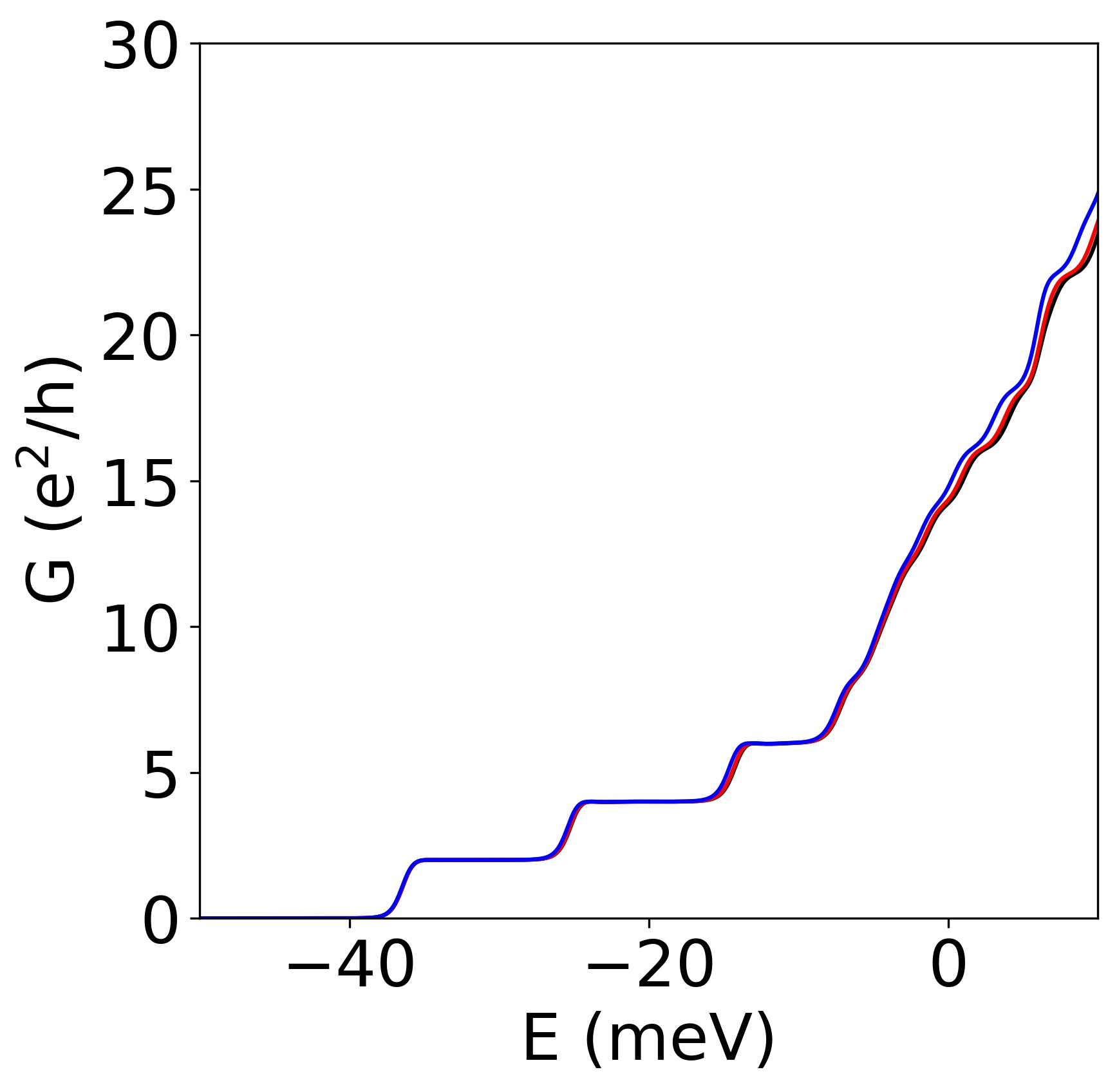} \caption{Conductance $G$ as a function of electron energy $E$ determined for a nanowire with a width of $W=78$~nm, using the ordinary TBA with a lattice size of $a$ (black), as well as within the scaled model with the scaling factor $s=2$ (red line) and $s=4$ (blue line).}
\label{fig:G}
\end{figure}
Figure~\ref{fig:Ek} presents the dispersion relations $E(k_x)$ determined for two values of the nanowire width: $W=15.6$~nm (a,b) and $W=78$~nm (c,d) and the scaling factor (a,c) $s=2$ and (c,d) $s=4$, respectively. For comparison, results from the ordinary TBA, defined on the lattice with size $a$, are plotted in black. Due to the tetragonal potential, the low-energy spectrum of the nanowire is primarily determined by the $d_{xy}$ bands which is particularly pronounced for a wide nanowire. At higher energies, the heavier $d_{xz/yz}$ bands contribute to the electronic spectrum, resulting in states that are a superposition of the light $d_{xy}$ and heavy $d_{xz/yz}$ bands.
\begin{figure*}[!t]
 \begin{tabular}{llll}
 \includegraphics[width=0.465\columnwidth]{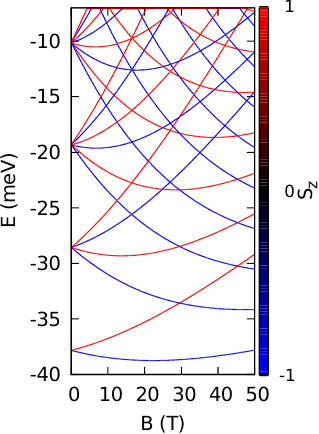}  \put(-75,30){(a)} &%
  \includegraphics[width=0.4\columnwidth]{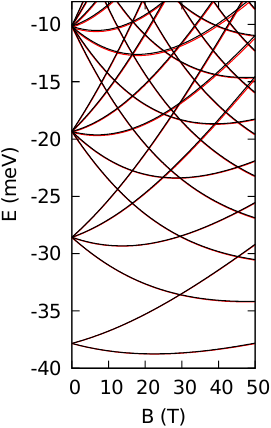}  \put(-60,30){(b)} &%
    \includegraphics[width=0.4\columnwidth]{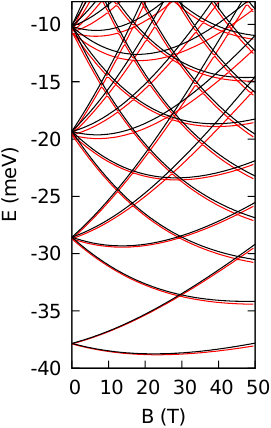}  \put(-60,30){(c)} &%
        \includegraphics[width=0.4\columnwidth]{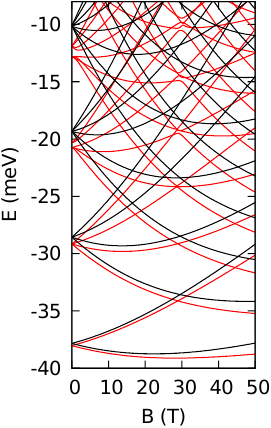}  \put(-60,30){(d)} \\%
 \end{tabular}
\caption{Single electron spectrum of QD for the energy confinement $\hbar \omega$=9.34 meV. (a) Results of the unscaled model with the color of the lines indicating the $z$-component of the electron spin. (b,c,d) Comparison of spectra evaluated using the unscaled model (black lines) and the scaled model with scaling factors of (a) $2$, (b) $4$, and (c) $8$. Results  
for the mesh size $62.8$~nm $\times$ $62.8$~nm corresponding to $161\times 161$ ions of Ti with 155526 spin-orbitals.}
 \label{fig:QD9}
\end{figure*}

\begin{figure*}[!t]
 \begin{tabular}{llll}
 \includegraphics[width=0.5\columnwidth]{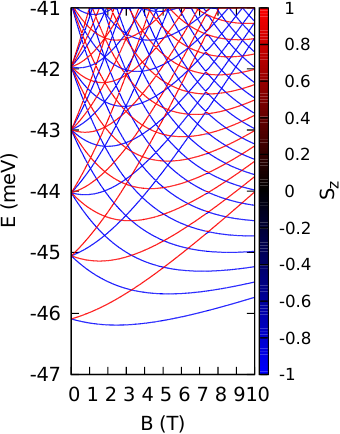}  \put(-85,30){(a)} &%
  \includegraphics[width=0.4\columnwidth]{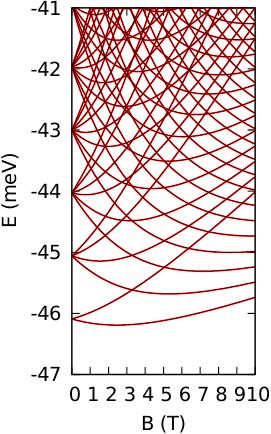}  \put(-60,30){(b)} &%
    \includegraphics[width=0.4\columnwidth]{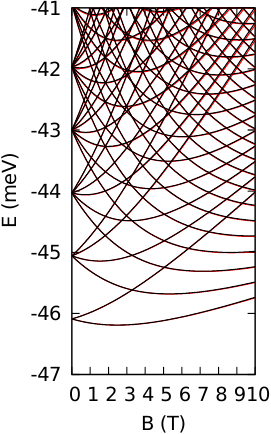}  \put(-60,30){(c)} &
     \includegraphics[width=0.4\columnwidth]{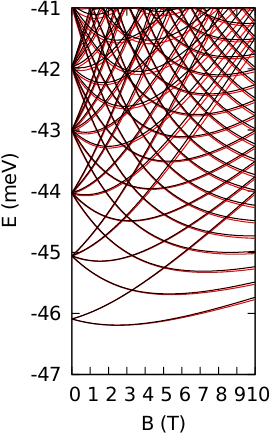}   \put(-60,30){(d)}\\%
 \end{tabular}
\caption{The same as in Fig.~\ref{fig:QD9} for the energy confinement $\hbar \omega$=1.04 meV. Results for the mesh size $156$~nm $\times$ $156$~nm corresponding to $401\times 401$ ions of Ti with 964806 spin-orbitals.
}
 \label{fig:QD1}
\end{figure*}

Note that the energy spectrum calculated on the grid with the scaling $s=2$ (a,c) quantitatively agrees with that obtained using the ordinary TBA. Increasing the scaling factor to $s=4$ makes this scaling procedure valid only for low energies, especially in the case of strong confinement [see Fig.\ref{fig:Ek}(b)]. Importantly, for larger widths [Fig.\ref{fig:Ek}(d)], as typically present in experiments, the scaled model with $s=4$ works quite well in the broad energy range.

For the wide nanowire, we have determined the conductance as a function of energy (see Fig.~\ref{fig:G}) using the scaled model with the scaling $s=2$ (red) and $s=4$ (blue). Fig.~\ref{fig:G} shows characteristics similar to those observed experimentally, with three well-pronounced steps that evolve into a quasi-linear dependence as the continuous part of the electronic spectrum, associated with the heavy $d_{xz/yz}$ bands, starts to be populated. Noticeably, the results of these calculations do not change significantly when using the scaled model with the scaling up to a factor of 4, which allows for simulations of much larger systems without a loss of accuracy.

\subsection{Electronic structure of confined systems}
\subsubsection{A single electron in a harmonic confinement}

Let us assume a harmonic oscillator confinement potential $V(r)=m\omega^2/2r^2$. The single-electron spectra obtained for the original TBA model are compared to results of the scaled version in Fig. \ref{fig:QD9} and \ref{fig:QD1} for the oscillator energy of $\hbar\omega=9.34$ meV and $\hbar\omega=1.04$ meV, respectively. For Fig. \ref{fig:QD9} we used a flake of 62.8 nm $\times$ 62.8 nm or 161 $\times$ 161 Ti ions with 155526 spin-orbitals in the unscaled model. For Fig. \ref{fig:QD1} the flake had a size of 156 nm $\times$ 156 nm, i.e. 401 $\times$ 401 Ti ions with 964806 spin-orbitals in the original model. Panels (a) of Figs. \ref{fig:QD9} and \ref{fig:QD1} show the spectra of the unscaled model with the color of the lines indicating the z-component of the electron spin. Panels (b), (c) and (d) display with the red lines the energy levels of the scaled model for the scaling factor of 2, 4 and 8. The energy levels coincide up to the thickness of the lines with the exact results for 
the scaling factor of 2 for $\hbar\omega=9.34$~meV (strong confinement), and 4 for $\hbar\omega=1.04$~meV (weak confinement).
\begin{table*}
\begin{tabular}{l|l|l|l|l|l|l}
$n$& $\delta E(s=2)$ ($\mu$eV) & $\delta E(s=4)$ ($\mu$eV)& $\delta E(s=8)$ ($\mu$eV) & $\delta r(s=2)$(pm) & $\delta r(s=4)$(pm)& $\delta r(s=8)$ (pm)  \\\hline
2 & 0.000 & -0.003 & -0.013 & 4.01 & 26.3 & 114.6\\
3 & 17.6 & 90.0 & 388.8 & 7.43 & 55.2 & 247.4\\
7 & 44.7 & 230.2 & 1086.0 & 10.0 & 96.1& 479.6 \\
13 & 82.9& 439.2& 2249.9 &4.71&160.1& 833.1
\end{tabular} 
\caption{The results for a single electron in the confined potential with $\hbar \omega$=9.34 meV and $B=0.5$T.
The first column gives the number of energy level. $\delta E(s)$ shows the difference
between the excitation energy of scaled and unscaled models $\delta E(s)\equiv \left[\Delta E_n(s=1)-\Delta E_1(s=1)\right]-\left[\Delta E_n(s)-\Delta E_1(s)\right]$.
 $\delta r(s)$ shows the difference
between the average size of the unscaled and scaled models $\delta r(s)\equiv \left[r_n(s=1)-r_1(s)\right]$, where $r_n=|\langle \Psi_n |(x^2+y^2)|\Psi_n\rangle|^{1/2}$.
}
\end{table*}

\begin{table*}
\begin{tabular}{l|l|l|l|l|l|l}
$n$& $\delta E(s=2)$ ($\mu$eV) & $\delta E(s=4)$ ($\mu$eV)& $\delta E(s=8)$ ($\mu$eV) & $\delta r(s=2)$(pm) & $\delta r(s=4)$(pm)& $\delta r(s=8)$ (pm)  \\\hline
2 & 0.000 & -0.001 & 1.84 & 1.73 & 8.68 & 36.58\\
3 & 0.186 & 0.928 & 5.76 & 3.44 & 17.23 & 72.82\\
7 & 0.446 & 1.374 & 11.28 & 5.42 & 27.20& 115.39 \\
13 & 0.782 & 3.916& 18.43 &7.66&38.44& 163.71
\end{tabular} 
\caption{Same as Table 1 only for $\hbar \omega$=1.04 meV.
}
\end{table*}

The precision of the scaled Hamiltonian can be read out from Tables I and II that display the differences in the energy and the mean radius of the electron density for  $\hbar\omega=9.34$ meV and $\hbar\omega=1.04$ meV. The results are taken for four energy levels: 2nd, 3rd, 7th and 13th -- each corresponding to subsequent energy level shells that are degenerate at $B=0$. The external magnetic field of $0.5$~T is taken for Tables I and II. The precision of the scaled Hamiltonian for the energy 
levels with the weaker confinement (Table II) is by two orders of magnitude better than for the stronger one (Table I). 
The precision for the radius of the charge density in both Tables is closer to each other, although the results for weaker confinement are still more accurate by a factor of 2 to 4. The exact radii obtained for the selected states are: $15.93$~nm, $22.52$~nm, $27.58$~nm and $31.84$~nm for Table I and $5.32$~nm, $7.51$~nm, $9.19$~nm and $10.57$~nm for Table II. 
\begin{figure}
\begin{tabular}{ll}
\includegraphics[width=0.4\columnwidth]{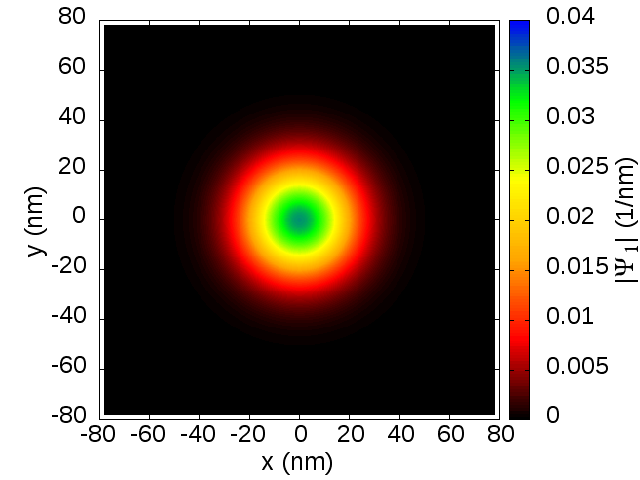}  \put(-10,10){(a)} &%
\includegraphics[width=0.4\columnwidth]{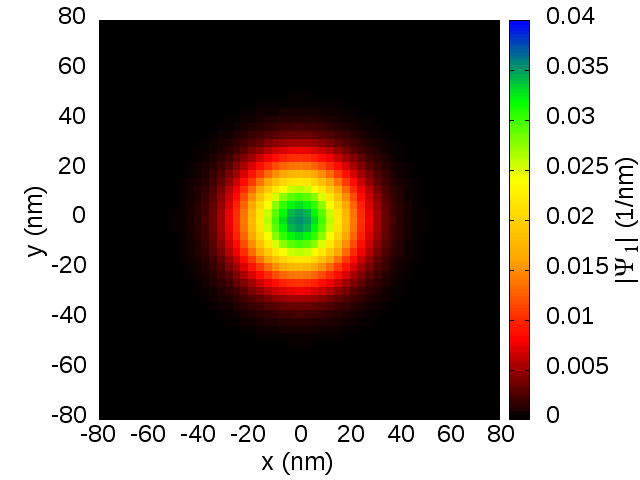}  \put(-10,10){(b)} \\
\includegraphics[width=0.4\columnwidth]{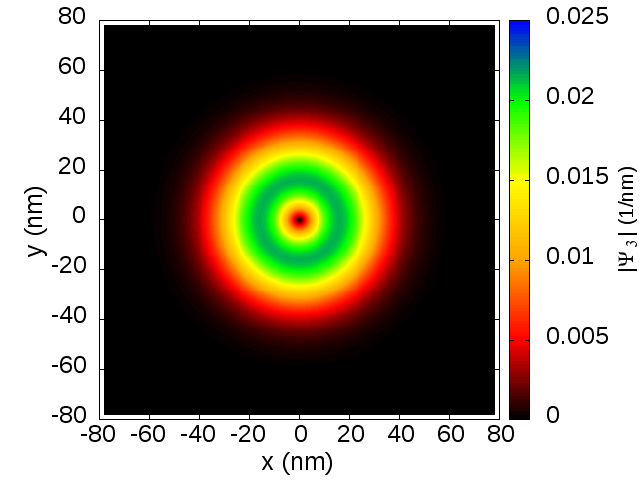}  \put(-10,10){(c)} &%
\includegraphics[width=0.4\columnwidth]{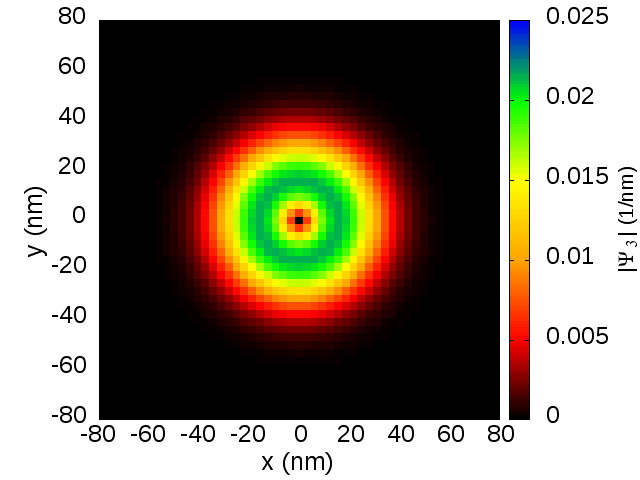}  \put(-10,10){(d)} \\
\includegraphics[width=0.4\columnwidth]{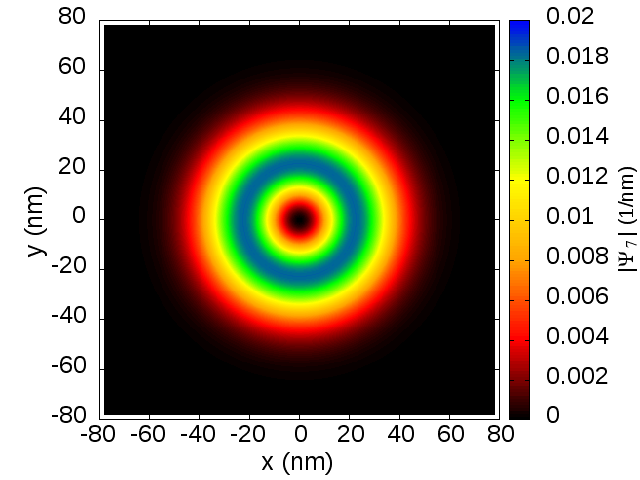}  \put(-10,10){(e)} &%
\includegraphics[width=0.4\columnwidth]{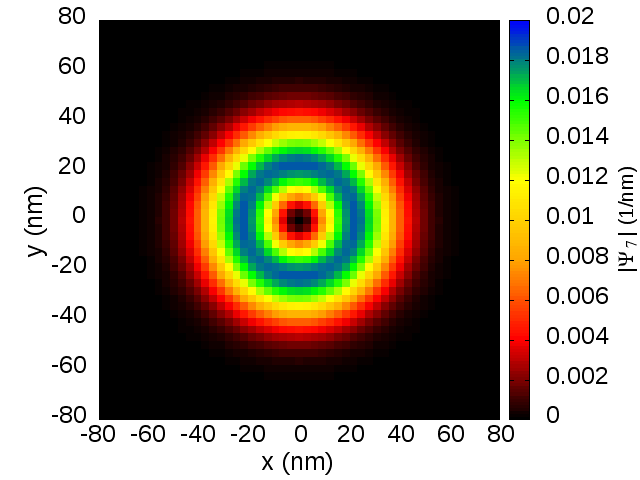}  \put(-10,10){(f)} \\
\includegraphics[width=0.4\columnwidth]{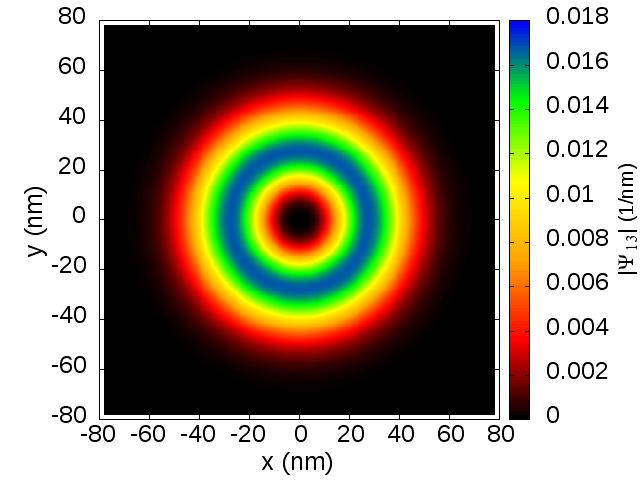}  \put(-10,10){(g)} &%
\includegraphics[width=0.4\columnwidth]{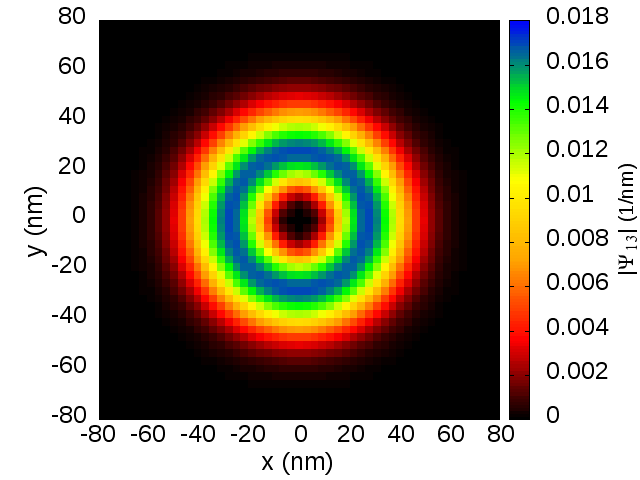}  \put(-10,10){(h)} \\
\end{tabular} 
\caption{An absolute value of the $d_{xy}$ spin down wave function component of the single-electron
ground state (a-b), 2nd excited state (c,d),  6th excited state (e,f) and 12th excited state (g,h) 
for $\hbar \omega$=1.04 meV and $B=0.5$T. The left column corresponds to the unscaled model
and the right one is obtained for the scaling factor $s=8$.}
\label{rosqd}
\end{figure}

Figure \ref{rosqd} shows the absolute value of the $d_{xy}$ spin-down component of the ground state, and 2nd, 6th, 12th, excited states for the unscaled Hamiltonian (left column) and the Hamiltonian with the scaling factor of 8 (right column). 
Although the results with the scaled Hamiltonian possess a considerably lower resolution, the overall character of the plots and the size of the wave function survive the scaling operation.
\begin{table}
\begin{tabular}{l|l|l|l|l|l|l}
$s$&  $x_{12}$(nm) &$x_{13}$(nm) \\ \hline
1 & 0.00590 & 0.3749 \\
2 & 0.00590 & 0.3748 \\
4 & 0.00589 & 0.3740 \\
8 & 0.00584 & 0.3708 \\
\end{tabular} 
\caption{Dipole matrix elements for the spin-flipping (second column) or spin-conserving 
(third column) transitions from the ground-state for $\hbar\omega=1.04$ meV.}
\end{table}
Additionally, in Table III we list the 
values of the dipole matrix elements for the transitions involving photons polarized in the $x$ direction from the ground-state to the second and third energy levels, i.e. with and without the spin-flip for the confinement energy of $\hbar\omega=1.04$ meV. The results of Table III indicate that
also the finer details of the wave functions are well reproduced by the scaled model, which should
therefore be useful for e.g. charge and spin dynamics of the driven systems~\cite{Szafran2024, Szafran2024_2}.

 Summarizing the results for a single electron in a harmonic oscillator confinement potential, we find
 that the scaled Hamiltonian provides results close to the exact ones for the weak confinement, i.e. for the case when the scaling is most needed.  Naturally, when the lateral confinement is strong and the wave function covers a smaller number of ions, the precision of the scaled model deteriorates for larger values of the applied scaling factor. For oscillator energy of 9.34 meV the scaling approach produces the energy spectra dependence on B that are identical up to a thickness of a line in the plot up to $s=2$. In comparison for the oscillator energy of 1.04 the separate lines for the unscaled and scaled method are resolved for $s$ as large as 8.

\subsubsection{An electron pair in a harmonic confinement}
Here, we consider a pair of electrons in a confinement potential using the Hamiltonian
\begin{equation}
\hat{H}_2=\hat{H}(1)+\hat{H}(2)+\frac{e^2}{4\pi\epsilon_0\epsilon r_{12}},
\end{equation}
with the dielectric constant of $\epsilon=100$~\cite{Szafran2024}. 
The two-electron eigenproblem is solved with the configuration interaction approach.
The single-electron Hamiltonian eigenstates are given by the basis of $3d$ spin-orbitals 
\begin{eqnarray}
\Psi_q(x,y,\sigma)&=&\sum_j a_j^q d_j(x,y,\sigma)\nonumber \\ &=&\sum_{r_j,o_j,s_j} a_j^q d_{r_j,o_j}(x,y)S_{s_j}(\sigma),
\end{eqnarray}
where $r_j$ is the position of $j$th ion, $o_j$ denotes one of the three $d$ spatial orbitals, and the spin $z$ component is denoted by $s_j$. 
The summing index $j$ stands for the triple indices $(r_j,o_j,s_j)$ and $d_{r_j,o_j}$ is one of the 3d orbitals localized on the ion position $r_j$ and $S$ stands for the spin-up or spin-down eigenstate.

Integration of the two-electron Hamiltonian matrix elements requires evaluation of the Coulomb integrals
\begin{eqnarray}
&&I_{q_1q_2q_3q_4}=\langle \Psi_{q_1}(1)\Psi_{q_2}(2)|\frac{1}{r_{12}}|\Psi_{q_3}(1)\Psi_{q_4}(2)\rangle \\ \nonumber &=& 
\sum_{j_1,j_2,j_3,j_4}(a_{j_1}^{q_1}a_{j_2}^{q_2})^*a_{j_3}^{q_3}a_{j_4}^{q_4}\langle d_{j_1}(1) d_{j_2}(2)|\frac{1}{r_{12}}|d_{j_3}(1) d_{j_4}(2)\rangle.
\end{eqnarray}
The integral over the spin-orbitals that appears in the sum is calculated based on the formula
\begin{eqnarray}
&&\langle d_{j_1}(1) d_{j_2}(2)|\frac{1}{r_{12}}|d_{j_3}(1) d_{j_4}(2)\rangle\nonumber = \\&&\delta(r_{j_1},r_{j_3})\delta(r_{j_2},r_{j_4}) \delta(s_{j_1},s_{j_3})\delta(s_{j_2},s_{j_4}) \times \nonumber \\&&
\bigg[ \left(1-\delta(r_{j_1},r_{j_2})\right)\frac{1}{|r_{j_1}-r_{j_2}|}\delta(o_{j_1},o_{j_3})\delta(o_{j_2},o_{j_4})
+ \nonumber \\&& \delta(r_{j_1},r_{j_2}) \varepsilon(o_{j_1},o_{j_2},o_{j_3},o_{j_4}) \bigg ],
\label{CI:integral}
\end{eqnarray}
where 
\begin{equation}
\varepsilon(o_{j_1},o_{j_2},o_{j_3},o_{j_4})=\langle d_{j_1}(1)d_{j_2}(2)|\frac{1}{r_{12}}|d_{j_3}(1)d_{j_4}(2) \rangle
\end{equation}
is single-ion integral,
 evaluated using the Monte-Carlo integration. For the orbitals  $d_{xy}=N \exp(-Z^*r/3)xy\rightarrow 1$, $d_{xz}=N \exp(-Z^*r/3)xz\rightarrow 2$, $d_{yz}=N \exp(-Z^*r/3)yz\rightarrow 3$, using the hydrogen-like orbitals with the effective atomic number $Z^*=3.65$ given by the Slater rules for the Ti orbitals and the normalization factor $N$ the non-zero values of the on-site integral are (in atomic units): $\varepsilon(i,i,i,i)=0.336$, $\varepsilon(i,j,i,j)=0.306$,
and $\varepsilon(i,j,j,i)=0.015$ (for $i\neq j$). The other single-ion integrals are zero. 
We use up to 50 lowest-energy single-electron states that produce 1225 Slater determinants for the two-electron problem. The scaling of the Ti ions mesh enters only via the single-electron Hamiltonian. We do not alter the single-ion integrals for $s>1$. 

\begin{figure*}
 \begin{tabular}{lll}
 \includegraphics[width=0.52\columnwidth]{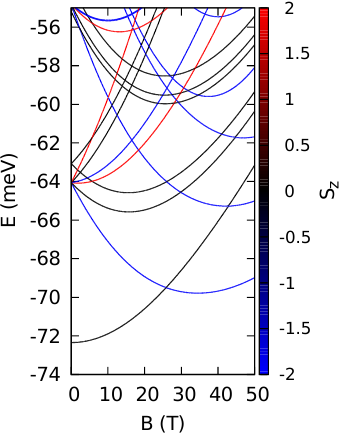}  \put(-90,30){(a)} &
  \includegraphics[width=0.41\columnwidth]{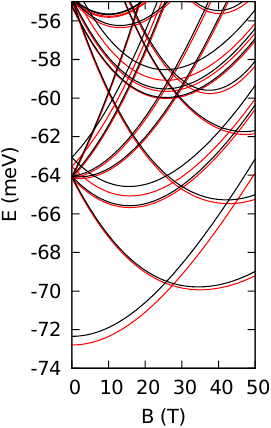}  \put(-60,30){(b)} &
    \includegraphics[width=0.41\columnwidth]{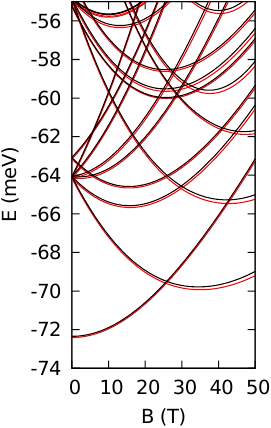}  \put(-60,30){(c)} %
 \end{tabular}
\caption{Electronic spectrum of two electrons in QD for the energy confinement $\hbar \omega$=9.34 meV. (a) Results of the exact model with the color of the lines indicating the $z$-component of the spin. (b) Comparison between the unscaled (black) and scaled (red) model with the scaling $s=2$.
(c) Same as (b) after scaling of the single-ion integral by $2.4$.
}
 \label{2e1k9}
\end{figure*}
\begin{figure*}
 \begin{tabular}{lll}
 \includegraphics[width=0.52\columnwidth]{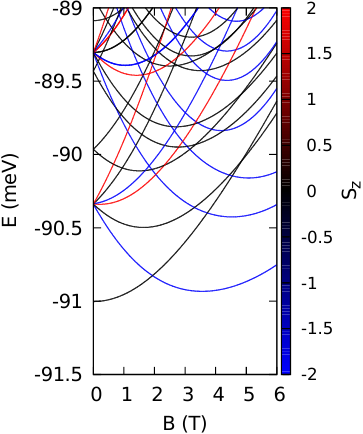}  \put(-90,30){(a)} &
  \includegraphics[width=0.41\columnwidth]{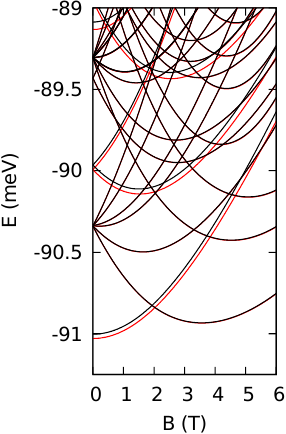}  \put(-60,30){(b)} &
    \includegraphics[width=0.41\columnwidth]{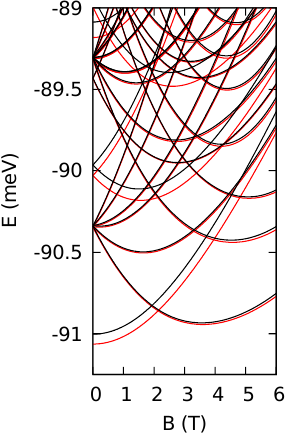}  \put(-60,30){(c)} %
 \end{tabular}
\caption{Electronic spectrum of two electrons in QD for the energy confinement $\hbar \omega$=1.04 meV. (a) Results of the exact model with the color of the lines indicating the $z$-component of the spin. (b,c) Comparison between the unscaled (black) and scaled (red) model with the scaling $s=2$ (b) and  $s=4$ (c).
}
 \label{2e1k1}
\end{figure*}

The two-electron spectrum for the unscaled Hamiltonian is given in Fig. \ref{2e1k9}(a) with 
the color of the lines corresponding to the z-component of the total spin. 
In Fig. \ref{2e1k9}(b) the exact results (black lines) are compared to the energy levels
obtained for the scaled Hamiltonian (red lines) with the scaling factor of 2. The two-electron states can be divided into two
groups corresponding to opposite symmetries with respect to the interchange of the spatial coordinates of the electrons: the ones that are symmetric and the ones that are antisymmetric with respect
to the operation of interchange of electron positions.
In Fig. \ref{2e1k9}(a) at $B=0$ the ground state is a spatially symmetric singlet. The first excited state is 6-fold degenerate state
which is spatially antisymmetric and is a counterpart of the spin-triplet but with extended degeneracy due to the orbital angular momentum.
The next energy level is two-fold degenerate at $B=0$ and both the states are again spatially symmetric spin singlets.
In Fig. \ref{2e1k9}(b) we notice that the energy of the symmetric ones (the ground-state for instance) are underestimated for the scaled Hamiltonian.
Namely, the energy difference between the exact eigenvalues and the eigenvalues of the scaled Hamiltonian with the scaling factor $s=2$
in Fig. \ref{2e1k9}(b) at $B=0$ are: $0.447$~meV for the ground-state, $0.082$~meV for the six-fold degenerate first excited state
and $0.465$~meV for the two-fold degenerate next excited state.  
The underestimation of the electron-electron interaction energy arises from the underestimation of the contribution of single-ion Coulomb integrals, which account for both electrons being located on the same ion. The scaled mesh contains fewer Ti ions resulting in the difference  in the overall on-site interaction energy.
The wave functions of the antisymmetric spatial part
have nodes for $\vec{r}_1=\vec{r}_2$. For these states the single-ion contributions to the 
Coulomb integrals are zero anyway, hence a much better precision in the energy estimate.
Fig. \ref{2e1k9}(c) presents the results for the single-ion integrals $\varepsilon$ increased by a factor of 2.4. The agreement between the spectra with a manipulated value of the single-ion integral is now much better. The factor of 2.4 has been adjusted to obtain this agreement. For varied potentials,
in shape or size, the relative contribution of the single-ion integrals to the two-center contributions varies and so is the value of the factor that produces the best fit of the scaled model to the exact results. However, for large quantum dots, i.e. where the scaling is most needed, the electrons also in the spatially symmetric states have enough space to avoid each other. Therefore, the two-electron spectra without any  adjustment of the $\varepsilon$ integrals reproduce the exact results much better for larger quantum dots.
Figure \ref{2e1k1} shows the results for $\hbar\omega=1.04$ meV. The agreement of the results with the scaling factor of 2 to the exact results  [Fig. \ref{2e1k1}(b)] is closer and the differences are well resolved only for the scaling factor of 4 [Fig. \ref{2e1k1}(c)].  The energy differences between the exact and scaled Hamiltonians for $s=2$ in Fig. \ref{2e1k1}(b) at $B=0$ are: $0.0266$~meV for the ground-state, $0.00175$~meV for the six-fold degenerate first excited state and $0.0284$~mwV for the two-fold degenerate next excited state.  For the scaling factor of 4 the energy differences are: 0.061 meV,
0.0086 meV and 0.0646 meV, respectively.

\begin{figure*}
 \begin{tabular}{llll}
 \includegraphics[width=0.49\columnwidth]{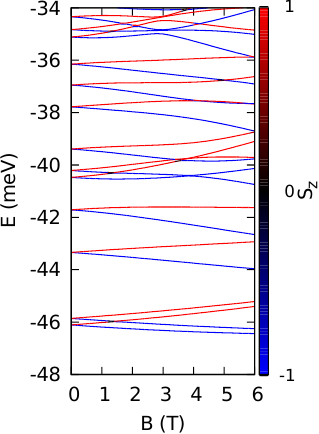}  \put(-85,30){(a)} &
 \includegraphics[width=0.4\columnwidth]{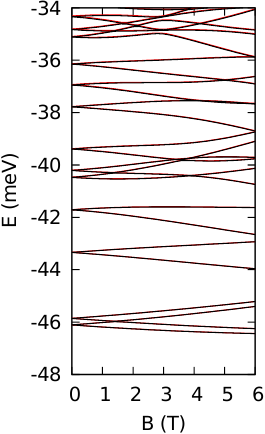}  \put(-60,30){(b)} &
  \includegraphics[width=0.4\columnwidth]{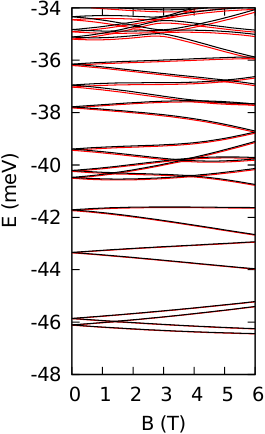}  \put(-60,30){(c)} &
  \includegraphics[width=0.4\columnwidth]{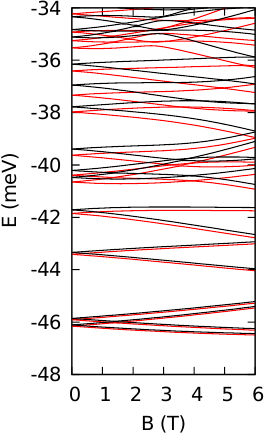}  \put(-60,30){(d)} \\
 \end{tabular}
\caption{Single electron spectrum for two Gaussian dots with $V_0=50$~meV, $R=20$~nm,  and $s_x=120a=46.8$~nm. (a) Results of the exact model with the color of the lines indicating the $z$-component of the spin. (b-d) Comparison between the unscaled (black) and scaled (red) model with the scaling $s=2$ (a), $s=4$ (b) and $s=8$ (d).}
 \label{fig:2k}
\end{figure*}

\begin{figure*}
 \begin{tabular}{llll}
 \includegraphics[width=0.51\columnwidth]{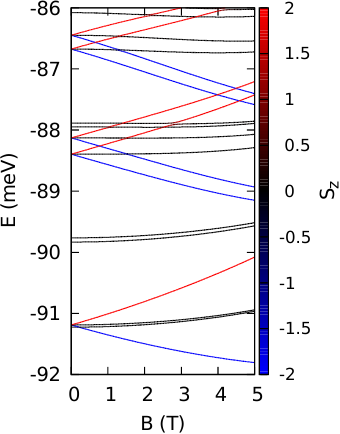}  \put(-95,27){(a)} &
 \includegraphics[width=0.4\columnwidth]{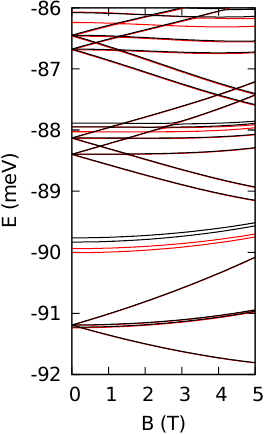}  \put(-70,27){(b)} &
  \includegraphics[width=0.4\columnwidth]{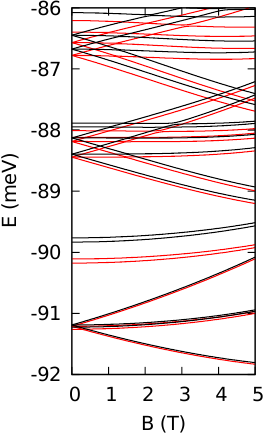}  \put(-70,27){(c)} 
 \end{tabular}
\caption{Same as Fig.~\ref{fig:2k} for pair of electrons and the scaling factor $s=2$ (b) and $s=4$ (c).}
 \label{fig:2k2e}
\end{figure*}

\subsubsection{Double quantum dot}
Let us replace the harmonic oscillator potential by a double quantum dot confinement \begin{eqnarray}V(x,y)&=&-V_0\exp\left(-\frac{1}{2}\left((x+\frac{s_x}{2})^2+y^2\right)/R^2\right) \nonumber \\&-&V_0\exp\left(-\frac{1}{2}\left((x-\frac{s_x}{2})^2+y^2\right)/R^2\right),\end{eqnarray}
where $V_0=50$ meV is the depth of the two Gaussian quantum dots \cite{SB,TC} with $R=20$ nm and $s_x$ is the spacing between the Gaussian centers. 
In Fig. \ref{fig:2k} we plot the single-electron spectrum with the unscaled [Fig. \ref{fig:2k}(a)] and scaled Hamiltonians [Fig. \ref{fig:2k}(b-d)]. 
The Hamiltonian results with the scaling factor of 2 are indistinguishable from the exact results, and only for larger values of the scaling parameter
the differences between the energy levels become visible. Therefore, the applicability of the scaling method for confined systems is not limited to the harmonic potential.

\begin{figure}
\begin{tabular}{ll}
 \includegraphics[width=0.5\columnwidth]{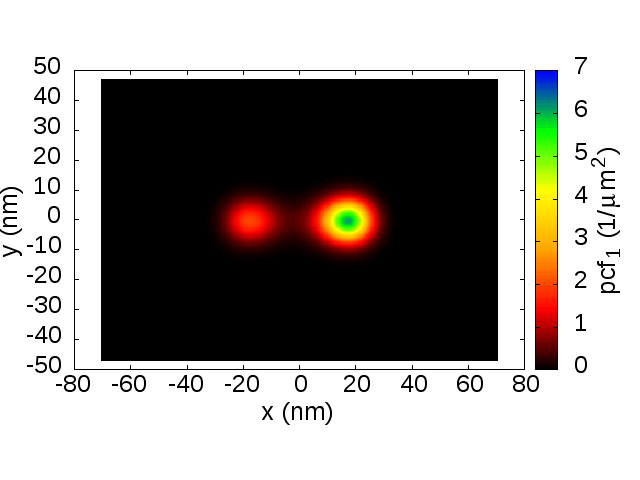}  \put(-10,10){(a)} &%
  \includegraphics[width=0.5\columnwidth]{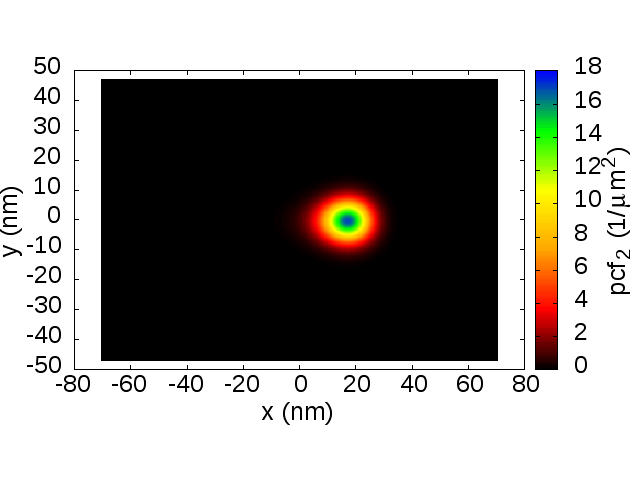}  \put(-10,10){(b)} \\
    \includegraphics[width=0.5\columnwidth]{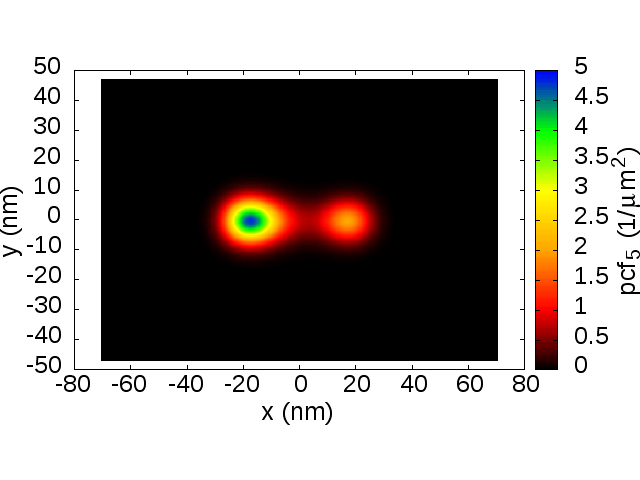}  \put(-10,10){(c)} &%
      \includegraphics[width=0.5\columnwidth]{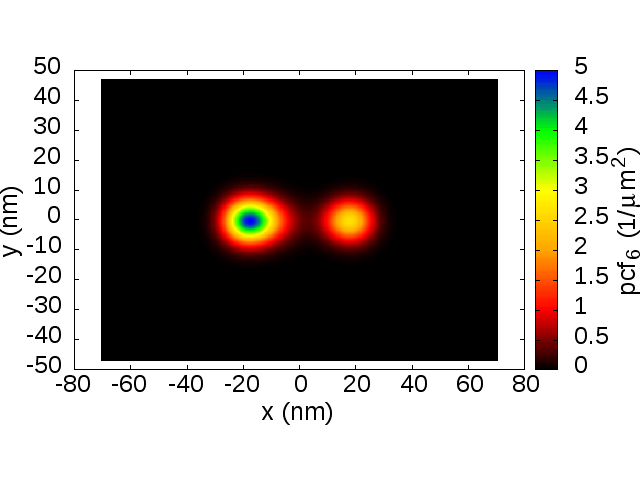}  \put(-10,10){(d)} \\
\end{tabular} 
\caption{Pair correlation function for position of a single electron fixed at the center of the left QD, at position ($-19.1$~nm,$0$),
for the ground-state singlet (a), the lowest-energy triplet (b), the lowest (c) and second (d) excited singlet states for $B=0$~T. The parameters of the double dot system as in Fig. \ref{fig:2k2e}.}  \label{pcf}
\end{figure}

The two-electron spectrum for the double quantum dot system is displayed in Fig. \ref{fig:2k2e}. The scaling approach
works very well for the lowest quadruple of energy levels: the lowest spin singlet and the lowest spin triplet. 
In these states the electrons avoid one another to minimize the Coulomb repulsion and stay mostly in opposite quantum dots. 
For these states, the contribution of the single-ion Coulomb integrals is small resulting in the high precision of the scaled Hamiltonian. The precision is much lower for the two-energy levels above the lowest quadruplet. 
Fig. \ref{pcf} shows the pair-correlation function, i.e. the probability of finding an electron when one of the electrons position is fixed. In Fig. \ref{pcf} the electron position is fixed near the center of the left quantum dot. 
In the ground-state spin singlet [Fig. \ref{pcf}(a)] the second electron can be found in the left dot, but the probability
of finding it in the right quantum dot is much larger. In the lowest spin-triplet state [Fig. \ref{pcf}(b)] the second electron
is with a near 100\% probability in the right quantum dot. A double occupation of the same dot is allowed in the spin-singlet state but not for the triplet state where it is forbidden by the Pauli exclusion. Fig. \ref{pcf}(c-d) show the results
for the two singlet states with the energy just above -90 meV. We can see that in both these excited states the electrons
tend to occupy the same dot, hence the contribution of the single-ion Coulomb integrals is large. As a result they are not well reproduced by the scaled Hamiltonian and a larger energy underestimation is found in Fig.~\ref{fig:2k2e}(c,d).
The energy for the lowest singlet for the scaled Hamiltonian differ from the exact ones at $B=0$ by $0.009561$~meV for the scaling factor of $2$ [Fig. \ref{fig:2k2e}(b)] and $0.03357$~meV for the scaling factor of $4$ [Fig. \ref{fig:2k2e}(c)].
The values for the lowest-energy triplet are $0.0043$~meV and $0.0217$~meV, respectively. 
Finally, for the second-energy singlet, for which the electrons tend to stay in the same dot the energy underestimates are $0.174$~meV and $0.343$~meV. 
Summarizing, for the two-electron states in a double quantum dot, the scaling method works with a high precision for the lowest-energy quadruple of energy levels including the spin singlet and the spin triplet that are most adequate for the spin-flip operations for quantum computing. 

\section{Summary}
We considered the scaled TBA approximation for the (001) LAO/STO 2DEG on a square lattice with a lattice constant $dx=s a$, defined by the scaling factor $s$. The results of the scaled model were compared to those of the exact one, defined on a lattice with the lattice constant $a=0.394$~nm.

First, we examined the applicability of the scaled model for simulating quantum transport, using as an example the transport through the QPC, which was recently demonstrated experimentally~\cite{Jouan2020}. Our analysis revealed the characteristic features of the conductance observed in the experiment and showed that the scaled model gives the same results as the exact one, with the accuracy increases for wide nanowires. In particular, for the nanowire width $W=78$~nm, the model with a scaling factor of up to $4$ yielded the same conductance characteristics as the unscaled approach.

Second, we analyzed the applicability of the scaled model for determining the electronic spectrum of single- and two-electron quantum dots. For the single-electron case, we found that the scaled Hamiltonian provides results close to the exact ones for weak confinement. When the lateral confinement is strong, and the wave function covers a smaller number of ions, the precision of the scaled model deteriorates for larger values of the scaling factor. Our calculations of two electrons revealed that the scaled model underestimates the contribution of single-ion Coulomb integrals, corresponding to both electrons at the same ion.  Interestingly, we found that the underestimation does not arise in double quantum dots, where the electrons tend to avoid each other to minimize Coulomb repulsion and primarily localize in opposite quantum dots. For these states, the contribution of single-ion Coulomb integrals is small, resulting in high precision of the scaled Hamiltonian.

In summary, our analysis demonstrates that the scaled model closely reproduces results of the exact one, especially for large systems (or weak confinement), within the realistic size range where the exact method becomes computationally demanding. This opens up new possibilities for simulating experimental setups with realistic dimensions.


\section*{Acknowledgments}
This work is financed by the Horizon Europe EIC Pathfinder under the grant IQARO number 101115190 titled "Spin-orbitronic quantum bits in reconfigurable 2D-oxides", we gratefully acknowledge Poland’s high-performance computing infrastructure PLGrid (HPC Center Cyfronet) for providing computer facilities and support within computational grant no. PLG/2024/017175 and no. PLG/2024/017043.


%

\end{document}